\documentclass[11pt,twoside]{article}
\usepackage{asp2010}

\resetcounters

\def\Ha{H$\alpha$ \,} 
\def\Hb{H$\beta$ \,} 

\def\ii{\,{\small II}} 
\def\iii{\,{\small III}}
\def\km{$\,\mbox{km}\,\mbox{s}^{-1}$}

\begin{document}

\title{Decoupled gas kinematics in isolated early-type disc galaxies}
\author{Katkov I.$^{1,2}$, Sil'chenko O.,$^{1}$ and Afanasiev V.$^3$
\affil{$^1$Sternberg Astronomical Institute, Lomonosov Moscow State University, Universitetskij pr., 13, 119992 Moscow, Russia}
\affil{$^2$Faculty of Physics, Lomonosov Moscow State University, Leninskie Gory 1, 119991, Moscow, Russia}
\affil{$^3$Special Astrophysical Observatory, Russian Academy of Sciences, Nizhnii Arkhyz, Karachai-Cherkessian Republic, 357147, Russia}
}

\begin{abstract}
We have studied a sample of completely isolated galaxies by means of long-slit
spectroscopy at the 6-m telescope. We have found that 7 of 12 ($58\pm14$\%) galaxies have
revealed a presence of large-scale ionized-gas component which angular momentum
is mostly differ from stellar one: 5 of 7 ($71\pm17$\%) show a visible
counterrottation. The diagnostic diagram demonstrates a wide range of gas
excitation mechanism. We have estimated the gas oxygen abundance in the cases where
excitation mechanism by young stars dominates and have found that ionized gas has a
subsolar metallicity. We concluded that cold-gas accretion from primordial
cosmological filaments is unlikely for these objects, while external accretion
from dwarf gas-rich satellites is more suitable.
\end{abstract}

\section{Introduction}
Evolutionary history of galaxies through cosmic time is thought to be
strongly controlled by a number of internal and external physical processes.
Consideration of isolated galaxies makes possible to constrain a diversity of
processes and to exclude some of them: gravitational tides, major mergers, ram
pressure stripping in hot intracluster/intragroup medium.

We have compiled our sample of isolated lenticular galaxies basing on the
approach recently developed by the team of Karachentsev, Makarov, and
co-authors.  They have developped a group-finding algorithm that takes into
account individual characteristics of galaxies in order to associate galaxies
in pairs, triplets and groups in the Local Supercluster and its surroundings
(see details in \citet{makarov_groups_2011} and references therein). We have
carried out spectroscopic observations of 12 strictly isolated galaxies with
isolation index $II>2.5$ and average value of $\overline{II}=12$: IC~875,
IC~1502, NGC~16, NGC~2350, NGC~3098, NGC~3248, NGC~6615, NGC~6654, NGC~6798,
NGC~7351, UGC~4551, UGC~9519. 

\section{Observations and data analysing}
We have observed spectroscopically the sample of lenticular galaxies using the
multimode focal reducer SCORPIO-2 \citep{scorpio2} mounted at the 6-m Russian
telescope. Long-slit spectra along major axes of galaxies were acquired with a
spectral resolution FWHM$\approx$4\AA\ in wavelength range from 3900 to 7000\AA.

\begin{figure*}
\centerline{
\includegraphics[width=0.33\textwidth]{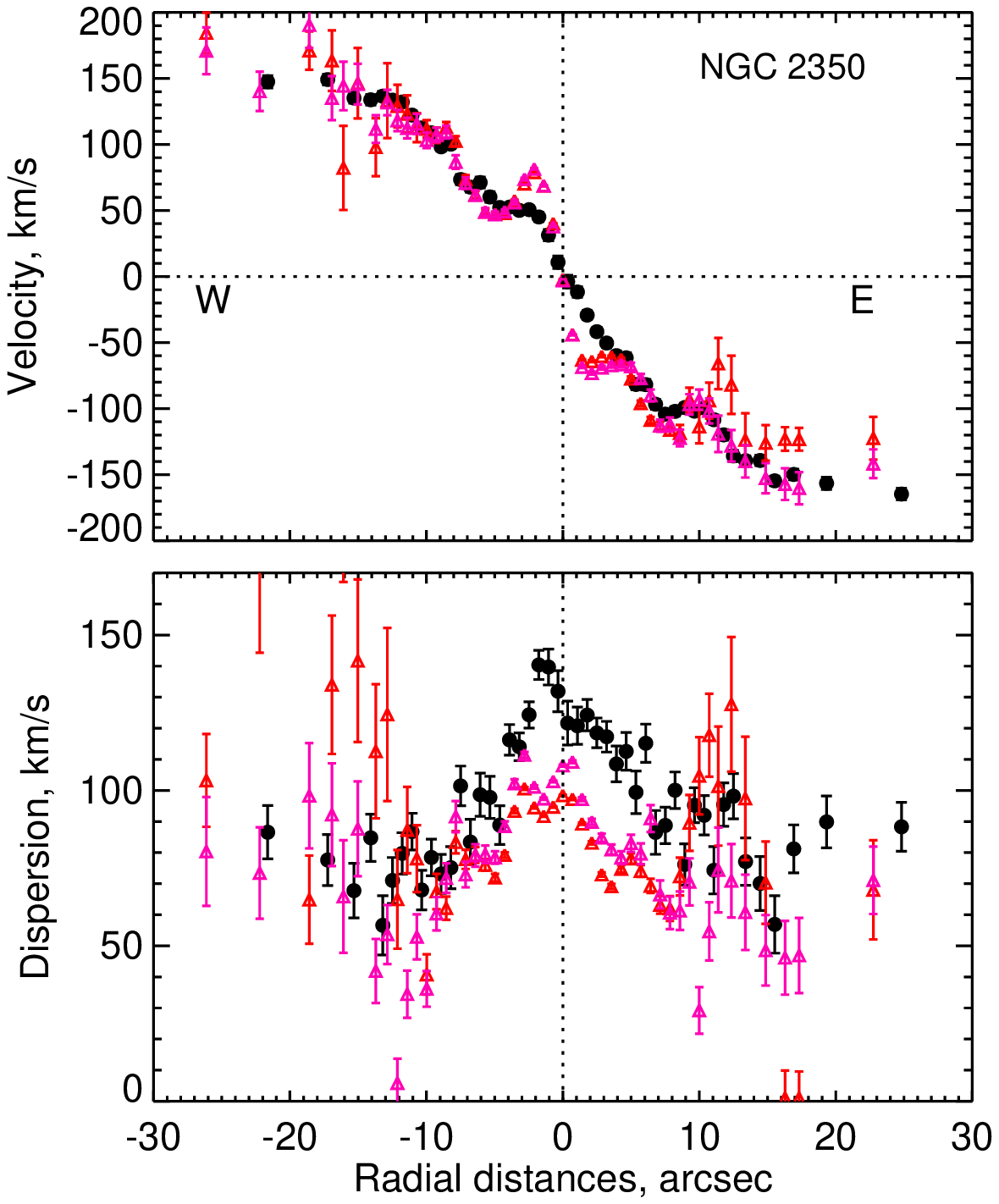}
\includegraphics[width=0.33\textwidth]{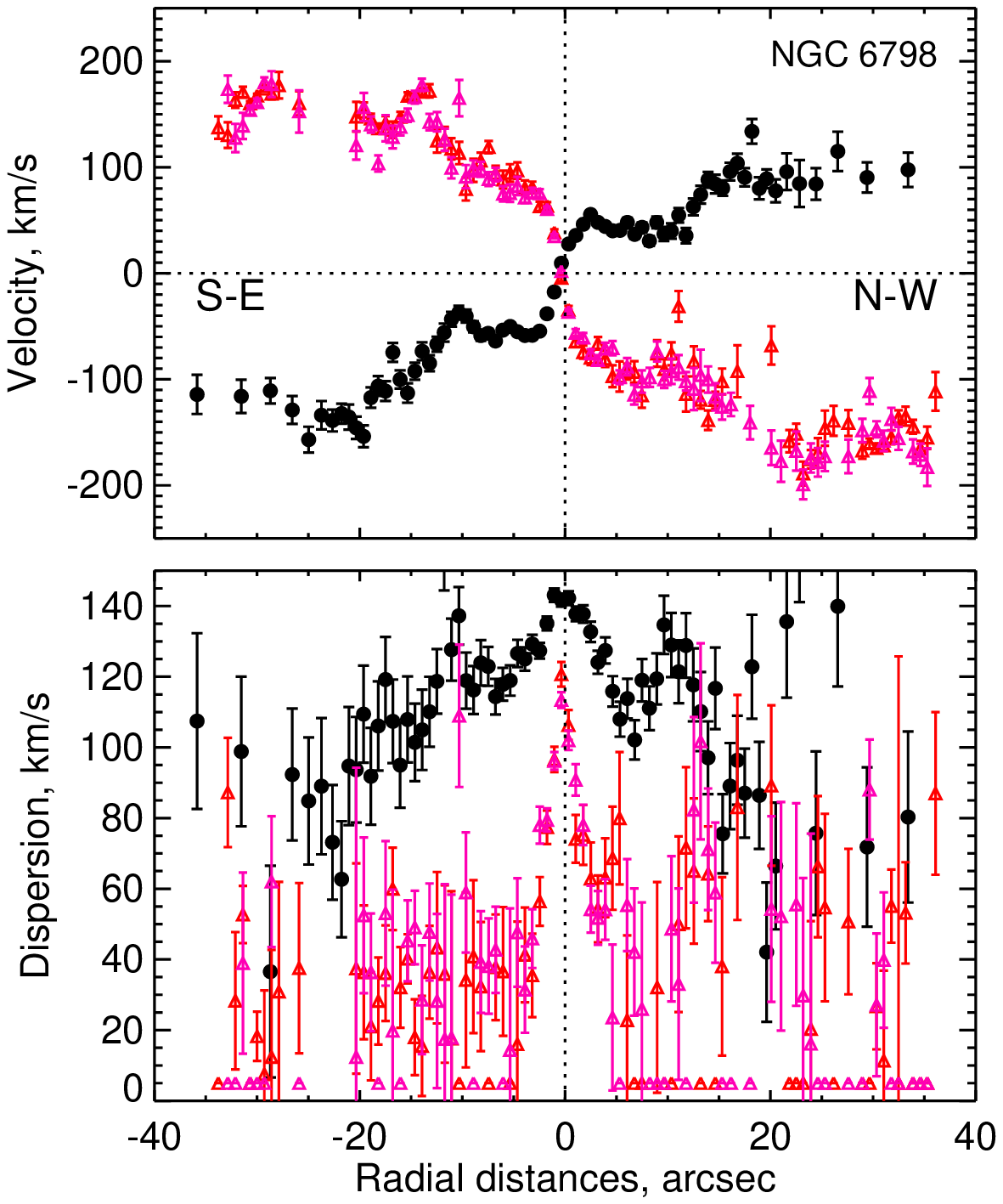}
\includegraphics[width=0.33\textwidth]{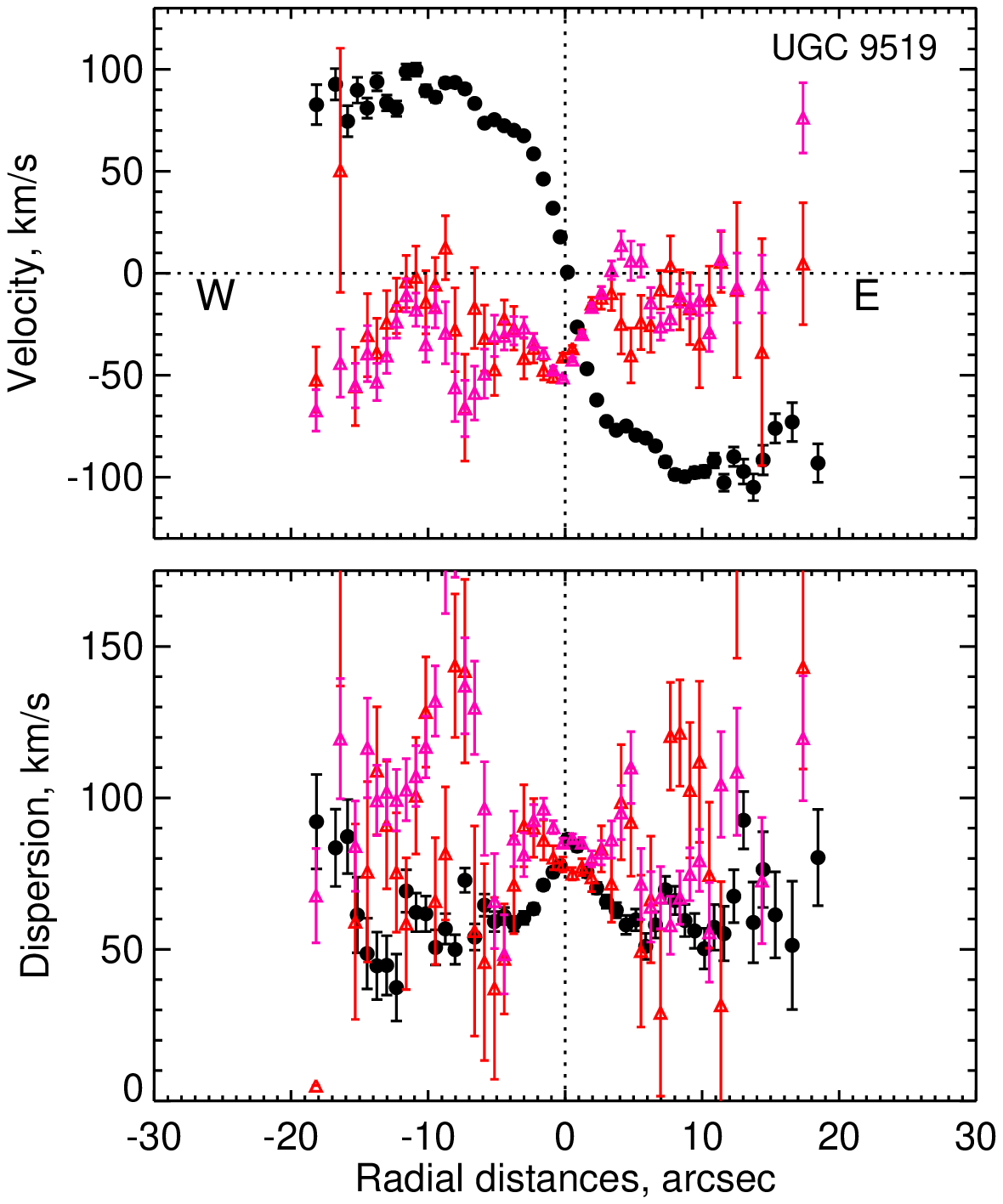}
}
\caption{Kinematical profiles of stars (black points) and ionized gas (color
points) of NGC~2350, NGC~6798, UGC~9519. \Ha measurements are shown by red colour,
[N\ii] -- magenta. The low velocity dispersions falling below the possibility of
being measured with our spectral resolution are indicated as 5 \km velocity
dispersions.} 
\label{fig_kin}
\end{figure*}

\begin{figure*}
\centerline{
\includegraphics[width=1.0\textwidth]{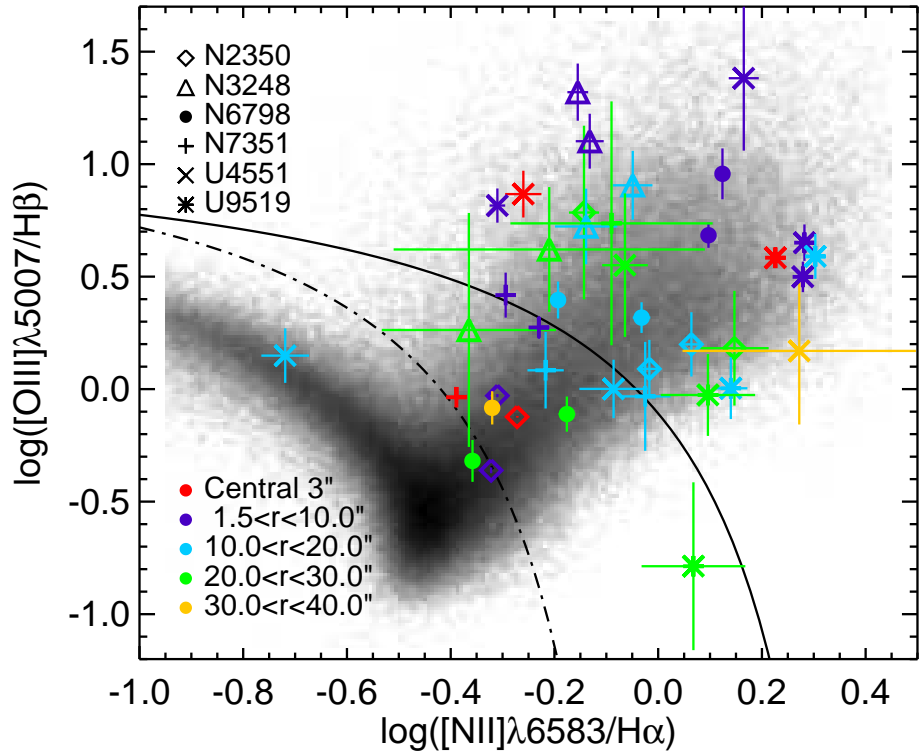}
}
\caption{Excitation diagnostic diagrams comparing the emission-line ratios:
[N\ii ]/\Ha vs. [O\iii]/\Hb . The colour of the points codes the distance
from the galaxy centers. The symbols correspond to different galaxies.    The
distribution of the measurements of the line ratios for galaxies from the SDSS
survey with high signal-to-noise ratios ($S/N>3$ in every line) are shown by
gray colour. The black curves, which separate the areas with the AGN/LINER
excitations (to the right) from areas with the star-formation-induced
excitation (to the left from these curves), are taken from
\citet{Kauffmann2003} (dash-dotted curve) and from \citet{Kewley2006} (solid
curve).} 
\label{fig_bpt}
\end{figure*}

%\subsection{Stellar kinematics}
To derive information about stellar and ionized gas kinematics we firstly
fitted the stellar absorption-line spectra by the PEGASE.HR high resolution stellar
population models \citep{pegasehr} convolved with a parametric line-of-sight
velocity distribution (LOSVD) by applying \textsc{NBursts} full spectral
fitting technique \citep{nbursts_a}. After that we subtracted the stellar absorption-line
model spectra from the observed spectrum to obtain pure emission-line spectrum, and then fitted 
it with Gaussians pre-convolved with instrumental resolution.  Kinematical profiles
for stars and ionized gas in a few most striking examples (NGC~2350, NGC~6798,
UGC~9519) are presented in Fig.~\ref{fig_kin}. To identify the dominant source
of the gas ionization, we plot our emission line measurements onto the
classical diagnostic BPT-diagram \citep{Baldwin1981bpt} (see
Fig.~\ref{fig_bpt}).

For the cases when the emission-line measurements reveal excitation dominated
by young stars, we determine oxygen abundance of ionized gas using NS-calibration
proposed by \citet{ns-calib}.

\section{Results and Discussion}
We have found that 7 of 12 ($58\pm14$\%) lenticular isolated galaxies studied by us
reveal a presence of extended ionized-gas discs which rotation is mostly
decoupled from the stellar kinematics: in 5 of 7 ($71\pm17$\%) -- NGC~3248,
NGC~6798, NGC~7351, UGC~4551, UGC~9519 -- the gas demonstrates visible 
counterrotation. Previously, \citet{atlas3d_10} studied
three galaxies of our list using their IFU data; they mentioned kinematical
misalignment of stars and ionized gas in these galaxies.

\citet{atlas3d_10} noted a dependence of gas kinematics in the early-type
galaxies (mostly S0s in the ATLAS-3D sample) on their environment: dense
environment provided tight coincidence between gas and star kinematics while in
more sparse environments the fraction of decoupled gaseous kinematics grew. Our
isolated lenticular galaxies represent an extreme point in this dependency, and
the fraction of decoupled gas kinematics exceeds 50 per cent.

The decoupled kinematics is a clear evidence for external origin of the gas
component.  Our galaxies are isolated so they cannot acquire their gas from
neighbours of comparable mass/luminosity; the sources of cold gas accretion may
be dwarf satellite merging or perhaps cosmological gas filaments.  Our
measurements of oxygen abundance for 3 objects indicate that gas metallicity is
subsolar, but not very low. NGC~6798 has a starforming ring at $r=\pm30$
arcsec where $12+\log O/H = 8.4-8.5$ dex. Oxygen abundance at the distance
$7^{\prime \prime}-8^{\prime \prime}$ in the NGC~7351 has value of 8.2 dex, 
while at the center 8.45 dex. Such gas metallicities are more consistent with 
the gas accretion from dwarf satellites where the gas could be enriched 
by metals than with accretion from cosmic filaments.

We probably have found some traces of dichotomy in the ionized-gas excitation
related to geometry of a gaseous system. When the gas is accreted smoothly in
the plane of a stellar disc, there are more possibilities to conserve its
coolness with following ignition of star formation (SF). The location of
emission-line ratio measurements in the SF region of the BPT diagram supposes
that the young stars radiation contributes there the dominant source of gas
ionization. Our observational data indicate that in the galaxies where the gas
is probably confined to the disc planes (NGC~2350, NGC~6798, NGC~7351, the very
outer part of NGC~6654) the excitation by young stars is preferable. Shocks or
post-AGB stars are the main agent of gas excitation in the remaining galaxies
(NGC~3248, UGC~4551, UGC~9519) whose velocity profiles possess asymmetries and
complex features which are probably resulted from gas motions in 
inclined planes.

%%%%%%%%%%%%%%%%%%%%%%%%%%%%%%%%%%%%%%%%%%%%%%%%%%%%%%%%%%%%%%%%%%%%%%%
%%%%%%%%%%%%%%%%%%%%%%%%%%%%%%%%%%%%%%%%%%%%%%%%%%%%%%%%%%%%%%%%%%%%%%%

%%%%%%%%%%%%%%%%%%%%%%%%%%%%%%%%%%%%%%%%%%%%%%%%%%%%%%%%%%%%%%%%%%%%%%%
%%%%%%%%%%%%%%%%%%%%%%%%%%%%%%%%%%%%%%%%%%%%%%%%%%%%%%%%%%%%%%%%%%%%%%%

\acknowledgements This work was supported by the RFBR grants 13-02-00059a,
12-02-00685a, 12-02-31452. I.K. is also grateful for the financial support of
the `Dynasty' Foundation.

%\bibliography{editor}
\bibliography{aspauthor}

\begin{thebibliography}{}
\expandafter\ifx\csname natexlab\endcsname\relax\def\natexlab#1{#1}\fi
\expandafter\ifx\csname url\endcsname\relax
  \def\url#1{\texttt{#1}}\fi
\expandafter\ifx\csname urlprefix\endcsname\relax\def\urlprefix{URL }\fi
\providecommand{\eprint}[2][]{\url{#2}}

\bibitem[{{Afanasiev} \& {Moiseev}(2011)}]{scorpio2}
{Afanasiev}, V.~L., \& {Moiseev}, A.~V. 2011, Baltic Astronomy, 20, 363

\bibitem[{{Baldwin} et~al.(1981){Baldwin}, {Phillips}, \&
  {Terlevich}}]{Baldwin1981bpt}
{Baldwin}, J.~A., {Phillips}, M.~M., \& {Terlevich}, R. 1981, \pasp, 93, 5

\bibitem[{{Chilingarian} et~al.(2007){Chilingarian}, {Prugniel}, {Sil'Chenko},
  \& {Koleva}}]{nbursts_a}
{Chilingarian}, I., {Prugniel}, P., {Sil'Chenko}, O., \& {Koleva}, M. 2007, in
  IAU Symposium, edited by A.~{Vazdekis}, \& R.~{Peletier}, vol. 241 of IAU
  Symposium, 175. \eprint{0709.3047}

\bibitem[{{Davis} et~al.(2011){Davis}, {Alatalo}, {Sarzi}, {Bureau}, {Young},
  {Blitz}, {Serra}, {Crocker}, {Krajnovi{\'c}}, {McDermid}, {Bois}, {Bournaud},
  {Cappellari}, {Davies}, {Duc}, {de Zeeuw}, {Emsellem}, {Khochfar},
  {Kuntschner}, {Lablanche}, {Morganti}, {Naab}, {Oosterloo}, {Scott}, \&
  {Weijmans}}]{atlas3d_10}
{Davis}, T.~A., {Alatalo}, K., {Sarzi}, M., {Bureau}, M., {Young}, L.~M.,
  {Blitz}, L., {Serra}, P., {Crocker}, A.~F., {Krajnovi{\'c}}, D., {McDermid},
  R.~M., {Bois}, M., {Bournaud}, F., {Cappellari}, M., {Davies}, R.~L., {Duc},
  P.-A., {de Zeeuw}, P.~T., {Emsellem}, E., {Khochfar}, S., {Kuntschner}, H.,
  {Lablanche}, P.-Y., {Morganti}, R., {Naab}, T., {Oosterloo}, T., {Scott}, N.,
  \& {Weijmans}, A.-M. 2011, \mnras, 417, 882. \eprint{1107.0002}

\bibitem[{{Kauffmann} et~al.(2003){Kauffmann}, {Heckman}, {Tremonti},
  {Brinchmann}, {Charlot}, {White}, {Ridgway}, {Brinkmann}, {Fukugita}, {Hall},
  {Ivezi{\'c}}, {Richards}, \& {Schneider}}]{Kauffmann2003}
{Kauffmann}, G., {Heckman}, T.~M., {Tremonti}, C., {Brinchmann}, J., {Charlot},
  S., {White}, S.~D.~M., {Ridgway}, S.~E., {Brinkmann}, J., {Fukugita}, M.,
  {Hall}, P.~B., {Ivezi{\'c}}, {\v Z}., {Richards}, G.~T., \& {Schneider},
  D.~P. 2003, \mnras, 346, 1055. \eprint{arXiv:astro-ph/0304239}

\bibitem[{{Kewley} et~al.(2006){Kewley}, {Groves}, {Kauffmann}, \&
  {Heckman}}]{Kewley2006}
{Kewley}, L.~J., {Groves}, B., {Kauffmann}, G., \& {Heckman}, T. 2006, \mnras,
  372, 961

\bibitem[{{Le Borgne} et~al.(2004){Le Borgne}, {Rocca-Volmerange}, {Prugniel},
  {Lan{\c c}on}, {Fioc}, \& {Soubiran}}]{pegasehr}
{Le Borgne}, D., {Rocca-Volmerange}, B., {Prugniel}, P., {Lan{\c c}on}, A.,
  {Fioc}, M., \& {Soubiran}, C. 2004, \aap, 425, 881.
  \eprint{arXiv:astro-ph/0408419}

\bibitem[{{Makarov} \& {Karachentsev}(2011)}]{makarov_groups_2011}
{Makarov}, D., \& {Karachentsev}, I. 2011, \mnras, 412, 2498.
  \eprint{1011.6277}

\bibitem[{{Pilyugin} \& {Mattsson}(2011)}]{ns-calib}
{Pilyugin}, L.~S., \& {Mattsson}, L. 2011, \mnras, 412, 1145.
  \eprint{1011.1431}

\end{thebibliography}

\end{document}